\documentclass{iopconfser}

\usepackage{graphicx}
\usepackage{dcolumn}
\usepackage{color}
\usepackage{bm}
\usepackage{amsmath,amssymb,graphicx}
\usepackage{epsfig}
\usepackage{enumerate}

\begin{document}

\title{Pseudo-Hermitian extensions of the harmonic and isotonic oscillators}

\author{Aritra Ghosh and Akash Sinha}

\affil{School of Basic Sciences, Indian Institute of Technology Bhubaneswar, Jatni, Khurda, Odisha 752050, India}

\email{ag34@iitbbs.ac.in}

\begin{abstract}
In this work, we describe certain pseudo-Hermitian extensions of the harmonic and isotonic oscillators, both of which are exactly-solvable models in quantum mechanics. By coupling the dynamics of a particle moving in a one-dimensional potential to an imaginary-valued gauge field, it is possible to obtain certain pseudo-Hermitian extensions of the original (Hermitian) problem. In particular, it is pointed out that the Swanson oscillator arises as such an extension of the quantum harmonic oscillator. For the pseudo-Hermitian extensions of the harmonic and isotonic oscillators, we explicitly solve for the wavefunctions in the position representation and also explore their intertwining relations.
\end{abstract}

	\section{Introduction}\label{introsec}
The recent years have experienced a considerable amount of interest in the study of non-Hermitian quantum systems \cite{NH,NH0,NH1,NH2,NH3,NH4}, and in particular, the ones that admit \(\mathcal{PT}\)-symmetry, wherein the Hamiltonian commutes with the combined action of the parity and time-reversal operators \cite{PT1,PT2,PTZ,PTZ1,Bagchi1,Bagchi2,BH,PT3,MUS,COR,Mich1,Mich2,example}. Such systems may admit real spectra despite being guided by non-Hermitian Hamiltonians. It should be remarked that non-Hermitian Hamiltonians are often encountered in optics and photonics \cite{MUS,phot1,phot2,WANG,EPO4} as well as in the study of thermal-atomic ensembles \cite{EPAMO}.


On one hand, a non-Hermitian system displays a non-unitary time evolution and may describe a quantum system that is interacting with some environment (could be a heat bath) \cite{NH0,NH4}, i.e., an open quantum system. This non-unitary time evolution turns out to be responsible for loss of quantum coherence (decoherence) \cite{deco2,decorev}. On the other hand, there is a class of non-Hermitian systems which yield a real spectrum and therefore may describe a unitary time evolution when viewed appropriately (see \cite{PTZ1} for some discussion on this). In the present study, we are interested in the latter class of non-Hermitian Hamiltonians. A Hamiltonian \(H\) is said to be pseudo-Hermitian if there exists some positive-definite Hermitian operator \(\rho\) such that \cite{mostafa,jones,jakubsky,mostafa2,das} (see \cite{znojiltd,fring,ORTIZ} for time-dependent generalizations)
\begin{equation}
 {H}^\dagger = \rho  {H} \rho^{-1},
\end{equation} which was shown to be the necessary condition for \( {H}\) to admit a real spectrum. Mostafazadeh \cite{mostafa,mostafa2} went on to show that under a similarity transformation
implemented by \(g = \sqrt{\rho}\), one has the following Dyson map:
\begin{equation}\label{Htohtransform}
h = g H g^{-1},
\end{equation} where \(h\) is a Hermitian Hamiltonian. Although \(H\) seems to dictate a non-unitary time evolution, the reality of the spectrum points towards an isolated (rather than open) system. Indeed the time evolution is unitary when viewed appropriately which involves the introduction of a metric operator \(\Theta = g^\dagger g\) that modifies the inner product as
\begin{equation}
(\psi,\Theta\phi) = (\psi,g^\dagger g\phi) = (g\psi, g\phi). 
\end{equation}
If \(\psi\) and \(\phi\) are the eigenfunctions of \(H\), one can now interpret \(g\psi\) and \(g\phi\) to be the eigenfunctions of its Hermitian equivalent, i.e., \(h\).


A prototypical pseudo-Hermitian quantum system is the Swanson oscillator which is described by the following Hamiltonian \cite{swanson1} (see also, Refs. \cite{jones,fring,ORTIZ,swanson2,swanson3,RAM,bagchiEP}):
\begin{equation}\label{SwansonH1111}
 {H}_{\rm Swanson} = \hbar \Omega_0 \bigg(  {a}^\dagger  {a} + \frac{1}{2}\bigg) + \alpha  {a}^2 + \beta ({ {a}^\dagger})^2,
\end{equation}
where \( {a}\) and \( {a}^\dagger\) are the lowering and raising operators, respectively, from the familiar harmonic-oscillator problem, i.e., they satisfy the algebra \([ {a}, {a}^\dagger] = 1\) with the other commutators vanishing identically. In Eq. (\ref{SwansonH1111}) above, \(\Omega_0 > 0\) and \(\alpha,\beta \in \mathbb{R}\). If \(\alpha \neq \beta\), then the Hamiltonian is non-Hermitian although it is \(\mathcal{PT}\)-symmetric as may be observed by noticing that under the action of \(\mathcal{PT}\), the lowering and raising operators transform as \( {a} \rightarrow -  {a}\) and \( {a}^\dagger \rightarrow - {a}^\dagger\), respectively. The system is pseudo-Hermitian and one can construct a Dyson map as \cite{jones} \(g ( {H}_{\rm Swanson}) g^{-1}  =  {h}\), where \( {h}\) is a Hermitian Hamiltonian (of the harmonic oscillator) and \(g\) is suitably chosen to dictate the above-mentioned transformation. This shows that the Swanson oscillator is closely related to its Hermitian counterpart, i.e., the harmonic oscillator. 


One of the aims of this paper is to demonstrate that the Swanson oscillator may be regarded as a harmonic oscillator coupled with an imaginary-valued gauge field. We shall analyze this problem in the position representation where we shall exactly solve for the wavefunctions and the spectrum of the oscillator. As may be expected, the wavefunctions are closely related to those of its Hermitian cousin, namely, the harmonic oscillator. In the same spirit, we will then study a pseudo-Hermitian extension of a harmonic oscillator with a centrifugal barrier, the so-called isotonic oscillator \cite{iso1,iso2}. Finally, their supersymmetric intertwining relations shall be exposed. 

\section{Quadratic Hamiltonians}
For our purposes, let us consider a quantum harmonic oscillator together with a non-Hermitian extension as
\begin{equation}\label{Hmodelcanquant}
 {H} = \frac{ {p}^2}{2m} + \frac{m\Omega^2  {x}^2}{2} + \frac{i \Lambda}{2} \big( {x}  {p} +  {p}  {x}\big),
\end{equation} where the part of the Hamiltonian which is proportional to \(\Lambda\) is non-Hermitian. The operators \( {x}\) and \( {p}\) satisfy \([ {x}, {p}] = i \hbar\). The Hamiltonian is \(\mathcal{PT}\)-symmetric, i.e., is invariant under the transformations \( {x} \rightarrow - {x}\), \( {p} \rightarrow  {p}\), and \(i \rightarrow -i\). Notice that the constant \(\Lambda\) in our model is a scalar and does not transform under \(\mathcal{PT}\). Thus, it may be expected that the Hamiltonian should support a real spectrum \cite{PT1}, and as we shall show later, indeed it does. This Hamiltonian can originate from a more general Hamiltonian that goes as
\begin{eqnarray}
 {H} &=& \frac{( {p} - A( {x}))^2}{2m} + V( {x}) \nonumber \\
&=& \frac{ {p}^2}{2m} +  \frac{ A( {x})^2}{2m} - \frac{( {p} A( {x}) + A( {x})  {p})}{2m}  + V( {x}), \label{HgencoupledtoA}
\end{eqnarray} 
wherein we choose\footnote{It should be pointed out that systems coupled with imaginary-valued gauge fields have generated significant research interest \cite{img,img1,img2,img3,img4}.} \(A( {x}) = - i \lambda  {x}\) and \(V( {x}) = \frac{m\omega^2  {x}^2}{2}\), for \(m > 0\) and \(\omega, \lambda \geq 0\). This gives Eq. (\ref{Hmodelcanquant}) if we identify
\begin{equation}\label{parameterdef}
\Lambda = \frac{\lambda}{m}, \quad \quad m \Omega^2 = m \omega^2 - \frac{\lambda^2 }{m},
\end{equation} which means\footnote{The case \(\Lambda = 0\) gives back the (Hermitian) harmonic oscillator. So we will assume that \(\Lambda > 0\).} \(\Lambda \geq 0\). Notice that in natural units, i.e., for \(\hbar =1\), both \(\Lambda\) and \(\Omega\) have dimensions of [length]\(^{-1}\). It is convenient to define a dimensionless parameter as
\begin{equation}\label{Omegargenericdef}
\Omega_r^2 = \bigg(\frac{\Omega}{\Lambda}\bigg)^2 = \bigg(\frac{\omega}{\Lambda}\bigg)^2 - 1.
\end{equation}
We may now have three situations listed as follows: 
\begin{enumerate}
\item If \(\omega \in (\Lambda,\infty)\), we have \(\Omega_r^2 > 0\). 
\item If \(\omega = \Lambda\), we have \(\Omega_r^2 = 0\). 
\item If \(\omega \in [0,\Lambda)\), we have \(\Omega_r^2 < 0\). 
\end{enumerate}
In the subsequent analysis, we will focus on the first two cases, i.e., where \(\Omega_r^2 \geq 0\). 

\subsection{Equivalence with Swanson oscillator}

\subsubsection{Scheme one}
Defining the raising and lowering operators in the usual way as
\begin{equation}\label{createdef1}
 {a}^\dagger = \frac{-i {p} + m \Omega  {x}}{\sqrt{2m \hbar \Omega}}, \quad \quad  {a} = \frac{i {p} + m \Omega  {x}}{\sqrt{2m \hbar \Omega}},
\end{equation}
where \([ {a}, {a}^\dagger]=1\), we have
\begin{equation}
 {p} = \sqrt{\frac{m\hbar \Omega}{2}} i( {a}^\dagger -  {a}), \quad \quad  {x} = \sqrt{\frac{\hbar}{2m\Omega}} ( {a}^\dagger +  {a}).
\end{equation}
In terms of these operators, the Hamiltonian operator given by Eq. (\ref{Hmodelcanquant}) reads as
\begin{equation}\label{Hswansonspecial}
 {H} = \hbar \Omega  {a}^\dagger  {a} + \frac{\hbar \Lambda }{2} ( {a}^2 - ({ {a}^\dagger})^2) + \frac{\hbar \Omega}{2},
\end{equation} which is a particular realization of the Swanson oscillator [Eq. (\ref{SwansonH1111})]; we have \(\Omega_0 = \Omega\), \(\alpha =  \frac{\hbar \Lambda}{2}\), and \(\beta = -  \frac{\hbar \Lambda}{2}\) for which Eq. (\ref{Hswansonspecial}) corresponds to Eq. (\ref{SwansonH1111}). In this case we have two independent parameters of the Hamiltonian given by Eq. (\ref{SwansonH1111}), i.e., \(\Omega_0\) and \(\alpha = (-\beta)\) which are determined by the parameters \( \Omega\) and \( \Lambda\); one must take \(\Omega > 0\) here.

\subsubsection{Scheme two}
Instead of defining the raising and lowering operators as in Eq. (\ref{createdef1}), let us define them as 
\begin{equation}
 {a}^\dagger = \frac{-i {p} +  {x}}{\sqrt{2}}, \quad \quad  {a} = \frac{i {p} +   {x}}{\sqrt{ 2}},
\end{equation} which satisfy the standard algebra \([ {a}, {a}^\dagger] = 1\) (setting \(\hbar = 1\)). This gives 
\begin{equation}
 {p} = \frac{i( {a}^\dagger -  {a})}{\sqrt{2}}, \quad \quad  {x} = \frac{( {a}^\dagger +  {a})}{\sqrt{2}},
\end{equation} and Eq. (\ref{Hmodelcanquant}) reads
\begin{eqnarray}
 {H} = \bigg(\frac{1}{2m} + \frac{m \Omega^2}{2}\bigg) \bigg( {a}^\dagger  {a} + \frac{1}{2} \bigg) + \bigg(-\frac{1}{4m} + \frac{m \Omega^2}{4} + \frac{\Lambda}{2}\bigg)  {a}^2 + \bigg(-\frac{1}{4m} + \frac{m \Omega^2}{4} - \frac{\Lambda}{2}\bigg) ( {a}^\dagger)^2 . \nonumber \\
 \label{Hswansonmoregeneral}
\end{eqnarray}
Since \(\hbar =1\), this Hamiltonian corresponds exactly to Eq. (\ref{SwansonH1111}) with 
\begin{equation}\label{parameterswansonmap}
\Omega_0 = \bigg(\frac{1}{2m} + \frac{m \Omega^2}{2}\bigg), \quad \alpha= \bigg(-\frac{1}{4m} + \frac{m \Omega^2}{4} + \frac{\Lambda}{2}\bigg), \quad \beta = \bigg(-\frac{1}{4m} + \frac{m \Omega^2}{4} - \frac{\Lambda}{2}\bigg).
\end{equation}
In this case we have three independent parameters of the Hamiltonian given by Eq. (\ref{SwansonH1111}), i.e., \(\Omega_0\), \(\alpha\), and \(\beta\); these are determined by the parameters \(m\), \(\Omega\), and \(\Lambda\) as is clear from Eq. (\ref{Hswansonmoregeneral}) or Eq. (\ref{parameterswansonmap}). Notice that for \(\Omega_r \geq 0\), one must have \(\Omega_0 \geq 0\). One can have \(\Omega_r \geq 0\), i.e., \(\Omega \geq 0\); if \(\Omega = 0\), then \(\alpha + \beta + \Omega_0 = 0\).

\subsection{Wavefunctions and spectrum}\label{simplesec}

For some generic values of \(\Omega\) and \(\Lambda\), the Hamiltonian [Eq. (\ref{Hmodelcanquant})] in the position representation reads as (for \(\hbar = 1\))
\begin{equation}\label{HOmeganonzerocase}
 {H} = -\frac{1}{2m} \frac{d^2}{dx^2} + \frac{ m \Omega^2 x^2}{2} + \frac{ \Lambda}{2} \bigg(2 x \frac{d}{dx} + 1\bigg).
\end{equation} 
Defining \(u = \sqrt{(m\Lambda/\sigma)} x\) with \(\sigma > 0\), one can show that Eq. (\ref{HOmeganonzerocase}) gives
\begin{equation}\label{HOmeganonzerocase1122}
 {H} = \frac{\Lambda}{2\sigma} \bigg[ - \frac{d^2}{du^2} + \Omega_r^2 \sigma^2 u^2 +  \sigma \bigg(2u \frac{d}{du} + 1\bigg)\bigg],
\end{equation}
which leads to a time-independent Schr\"odinger equation that goes as
\begin{equation}\label{abcde}
 \frac{d^2\psi(u)}{du^2} - \Omega_r^2 \sigma^2 u^2 \psi(u) - 2  \sigma u \frac{d\psi(u)}{du} + \sigma \bigg(\frac{2E }{\Lambda} - 1 \bigg) \psi(u) = 0,
\end{equation}
where \(\Omega_r^2\) is defined in Eq. (\ref{Omegargenericdef}). As can be observed from Eq. (\ref{HOmeganonzerocase1122}), \(\Lambda\) sets the energy scale corresponding to the Hamiltonian which is pseudo-Hermitian and can be mapped to that of a Hermitian oscillator as \(g  {H} g^{-1} =  {h}\), where
\begin{equation}
 {h} = \frac{\Lambda}{2 \sigma} \bigg[-\frac{d^2}{d u^2} + \sigma^2 u^2 (1 + \Omega_r^2) \bigg], \quad \quad g = e^{-\frac{\sigma u^2}{2}}.
\end{equation}
The form of \(g\) has been obtained by taking the ansatz \(g = e^{-({\rm constant}) u^2}\) and then by fixing the constant in the exponential by demanding that \(h\) is Hermitian. Following \cite{das}, the inner product between the eigenfunctions of the Hamiltonian given by Eq. (\ref{HOmeganonzerocase1122}) will be determined by the metric \(\Theta = g^\dagger g = e^{-\sigma u^2} = e^{-m\Lambda x^2}\).


Let us take \(\psi(u) = e^{\alpha_0 u^2} \phi(u)\) which gives
\begin{equation}\label{abcde1phi}
 \frac{d^2\phi(u)}{du^2}  + (4 \alpha_0 -2 \sigma) u \frac{d\phi(u)}{du} + (4 \alpha_0^2 -  4 \sigma \alpha_0-\Omega_r^2 \sigma^2) u^2 \phi(u) + \bigg[\sigma \bigg(\frac{2E }{\Lambda} - 1 \bigg) + 2\alpha_0 \bigg] \phi(u) = 0.
\end{equation}
This becomes a Hermite equation if the parameters satisfy the following conditions:
\begin{eqnarray}
4 \alpha_0 -2 \sigma &=& - 2, \label{gencondsigmaalpha} \\
4 \alpha_0^2 -  4 \sigma \alpha_0-\Omega_r^2 \sigma^2 &=& 0, \label{equationforalpha0} \\
\sigma \bigg(\frac{2E }{\Lambda} - 1 \bigg) + 2\alpha_0 &=& 2n, \quad n = 0,1,2,\cdots. \label{Econditiongeneral}
\end{eqnarray}
Eq. (\ref{equationforalpha0}) gives 
\begin{equation}
\alpha_0 = \frac{\sigma}{2} (1 \pm \sqrt{1 + \Omega_r^2}),
\end{equation} wherein we will pick the one with the negative sign as we desire \(\alpha_0\) to be negative so that the factor \(e^{\alpha_0 u^2}\) falls off for \(u \rightarrow \pm \infty\). 
Combining Eqs. (\ref{gencondsigmaalpha}) and (\ref{equationforalpha0}), it is found that \(\Omega_r\) must be given by 
\begin{equation}\label{Omegarsigmacondition}
\Omega_r = \sqrt{\frac{1}{\sigma^2} - 1},
\end{equation} and demanding \(\Omega_r^2 \geq 0\) implies that \(0 < \sigma \leq 1\); the case with \(\sigma = 1\) gives \(\Omega_r = 0\). Finally, combining Eqs. (\ref{gencondsigmaalpha}) and (\ref{Econditiongeneral}), the spectrum turns out to be  
\begin{equation}\label{spectrumgeneral}
E_n = \frac{\Lambda}{2 \sigma}(2n + 1),
\end{equation} which is equispaced and coincides with that of the harmonic oscillator since \(\Lambda =\sigma \omega\) from Eqs. (\ref{Omegargenericdef}) and (\ref{Omegarsigmacondition}). The wavefunctions go as (up to normalization factors) \(\psi_n(x) \sim e^{\frac{(\sigma-1)m\Lambda}{2\sigma} x^2}  H_n(\sqrt{m\Lambda/\sigma} x)\). The inner product is defined using the metric \(\Theta = g^\dagger g = e^{-m \Lambda x^2}\), which, upon using the well-known orthogonality property of the Hermite polynomials, i.e., \(\int_{-\infty}^\infty  H_{n_1}(\xi) H_{n_2}(\xi) e^{-\xi^2} d\xi = \sqrt{\pi} 2^{n_1} n_1! \delta_{n_1,n_2}\), gives \((\psi_{n_1},\Theta \psi_{n_2}) = 0\) for \(n_1\neq n_2\). Using the orthogonality property, we may normalize the wavefunctions to write 
\begin{equation}\label{wavefunctions11111111}
\psi_n(x) = \bigg(\frac{m\Lambda}{\sigma \pi}\bigg)^{1/4} \frac{e^{\frac{(\sigma-1)m\Lambda}{2\sigma} x^2} }{2^{n/2} \sqrt{n!}} H_n \bigg(\sqrt{\frac{m\Lambda}{\sigma}}x\bigg).
\end{equation}
Thus, we have completely solved the problem; the wavefunctions are given by Eq. (\ref{wavefunctions11111111}) while the corresponding spectrum is given by Eq. (\ref{spectrumgeneral}).

\section{Pseudo-Hermitian extension of the isotonic oscillator}
The isotonic oscillator \cite{iso1,iso2} is a close cousin of the harmonic oscillator. Although it admits a nonlinear Newton's equation of motion, it is isoperiodic with the harmonic oscillator \cite{CV,ISO4} and the corresponding quantum-mechanical problem admits an equispaced spectrum \cite{iso1,iso2}. The isotonic oscillator has also found applications in the theory of coherent states \cite{iso2,ISO6}. The potential describing the dynamics may be taken to be
\begin{equation}
V(x) = V_0 \bigg(\frac{x}{x_0} - \frac{x_0}{x}\bigg)^2, \quad \quad x>0,
\end{equation}  where \(x_0 > 0\) is a suitable constant with dimensions of length and \(V_0 > 0\) is a constant with dimensions of energy (for \(\hbar = 1\), it has dimensions of [length]\(^{-1}\)). As a generalization of this problem in the spirit of the ongoing analysis, let us consider the following generalized quantum Hamiltonian in the position representation: 
\begin{equation}\label{Hisogen}
 {H} =  -\frac{1}{2m}\frac{d^2}{dx^2} + V_0 \bigg(\frac{x}{x_0} - \frac{x_0}{x}\bigg)^2 + \frac{ \Lambda}{2} \bigg(2 x \frac{d}{dx} + 1\bigg). 
\end{equation}
Putting \(\xi = x/x_0\), the system has been exactly solved for the case \(\Lambda = 0\) by taking an ansatz that goes as \(\psi(\xi) = \xi^\nu e^{-\frac{\eta \xi^2}{4} } \phi(\xi)\), wherein \(\nu , \eta >0\) and it is found that \(\phi(\xi)\) satisfies Kummer's differential equation for the confluent hypergeometric function \cite{iso1,iso2} which can be related to the Laguerre polynomials \cite{PTZ,lagu}.

In order to address the quantum mechanics dictated by the non-Hermitian Hamiltonian given by Eq. (\ref{Hisogen}), let us first rewrite the Hamiltonian in terms of the variable \(\xi = x/x_0\) which gives
\begin{equation}\label{Hisogen1}
 {H} =  V_0 \bigg[ -\frac{4}{\eta^2}\frac{d^2}{d\xi^2} + \bigg(\xi - \frac{1}{\xi}\bigg)^2 + \frac{\Lambda_0}{2} \bigg(2 \xi \frac{d}{d\xi} + 1\bigg) \bigg],
\end{equation}
where \(\Lambda_0 = \Lambda/V_0\) and \(\eta^2 = 8 m V_0 x_0^2\). The Hamiltonian is pseudo-Hermitian and one can construct a Dyson map as \( {h} = g {H} g^{-1}\), for
\begin{equation}\label{hisoHermitian}
 {h} = - \frac{4 V_0}{\eta^2}\frac{d^2}{d\xi^2} +V_0 \left(\xi -\frac{1}{\xi }\right)^2 + \frac{1}{16} V_0 \eta ^2 \Lambda _0^2 \xi
^2, \quad \quad g = e^{-\frac{\Lambda_0 \eta^2 }{16}\xi^2},
\end{equation} where \({h}\) is Hermitian and the form of \(g\) has been obtained by taking the ansatz \(g = e^{-({\rm constant}) \xi^2}\) and then by fixing the constant in the exponential by demanding that \(h\) is Hermitian. Corresponding to Eq. (\ref{Hisogen1}), the time-independent Schr\"odinger equation reads
\begin{equation}
\frac{d^2\psi(\xi)}{d\xi^2} -  \frac{\eta^2}{4} \bigg(\xi - \frac{1}{\xi}\bigg)^2 \psi(\xi) - \frac{\Lambda_0 \eta^2}{4} \xi \frac{d \psi(\xi)}{d\xi} + \frac{\eta^2}{4} \bigg( \frac{E}{V_0} - \frac{\Lambda_0}{2}\bigg)\psi(\xi) = 0. 
\end{equation}
Let us define \(y = \eta \xi^2/2\) such that the time-independent Schr\"odinger equation reads as 
\begin{equation}
y\frac{d^2\psi(y)}{dy^2}  + \frac{1}{2} \bigg(1- \frac{\Lambda_0 \eta}{2} y\bigg) \frac{d \psi(y)}{dy} -  \frac{\eta}{4} \bigg(\frac{y}{\eta} + \frac{\eta}{4y} - 1\bigg) \psi(y) + \frac{\eta}{8} \bigg( \frac{E}{V_0} - \frac{\Lambda_0}{2}\bigg)\psi(y) = 0. 
\end{equation}
We now take an ansatz for the wavefunction which goes as \(\psi(y) = e^{- \alpha_0 y} y^{\beta_0} \phi(y)\), for some suitable real and positive parameters \(\alpha_0\) and \(\beta_0\). The function \(\phi(y)\) must satisfy the following equation:
\begin{equation}\label{eqnphiisotonic}
X(y)  \frac{d^2 \phi(y)}{dy^2} + Y(y) \frac{d\phi(y)}{dy} + Z(y) = 0,
\end{equation}
where 
\begin{eqnarray}
 X(y) &=& y, \nonumber \\
Y(y) &=& -2 \alpha_0 y + 2 \beta_0 + \frac{1}{2} \bigg(1- \frac{\Lambda_0 \eta}{2} y\bigg),   \nonumber \\
 Z(y) &=& \alpha_0^2 y - 2 \alpha_0 \beta_0 + \frac{\beta_0(\beta_0 - 1)}{y} + \frac{1}{2} \bigg(1- \frac{\Lambda_0 \eta}{2} y\bigg) \bigg(-\alpha_0 + \frac{\beta_0}{y}\bigg) \nonumber \\
&&- \frac{\eta}{4} \bigg(\frac{y}{\eta} + \frac{\eta}{4y} - 1\bigg) + \frac{\eta}{8} \bigg( \frac{E}{V_0} - \frac{\Lambda_0}{2}\bigg). \nonumber
\end{eqnarray}
The above-mentioned equation can be solved to yield the Laguerre polynomials as \cite{lagu}
\begin{equation}\label{phisollagu}
\phi_n (y) = L_{n}^{a}\left(\frac{1}{4} y
\sqrt{\eta^2 \Lambda_0^2+16}\right), \quad a = \frac{\sqrt{\eta ^2+1}}{2}, \quad n = \frac{1}{4} \left(\frac{2
(2 V_0+E_n) \eta }{V_0 \sqrt{\eta^2 \Lambda_0^2+16}}-\sqrt{\eta
^2+1}-2\right),
\end{equation}
where \(n = 0,1,2,\cdots\), and this indicates that the spectrum reads as
\begin{equation}
E_n =\frac{ \left(2
V_0 \sqrt{\eta^2 \Lambda_0^2+16}\right)}{\eta } \left(n+\frac{1}{2}+\frac{\sqrt{\eta ^2+1}}{4}-\frac{\eta }{\sqrt{\eta ^2 \Lambda_0
^2+16}}\right),
\end{equation}
which is equispaced. In obtaining Eq. (\ref{phisollagu}), the parameters \(\alpha_0\) and \(\beta_0\) turn out to be 
\begin{equation}
\alpha_0 = \frac{1}{8} \bigg(\sqrt{\eta ^2 \Lambda_0^2+16}-\eta  \Lambda_0 \bigg), \quad \quad \beta_0 = \frac{1}{4} \bigg(\sqrt{\eta ^2+1}+1\bigg),
\end{equation} so that the wavefunctions (up to normalization factors) read
\begin{equation}
\psi_n(\xi) \sim e^{- \frac{\eta \big(\sqrt{\eta ^2 \Lambda_0^2+16}-\eta  \Lambda_0 \big) \xi^2}{16}} \xi^{\frac{1}{2} \big(\sqrt{\eta ^2+1}+1\big)} L_{n}^{\frac{\sqrt{\eta ^2+1}}{2}}\left(\frac{\eta \xi^2}{8}
\sqrt{\eta^2 \Lambda_0^2+16}\right).
\end{equation}
The wavefunctions are consistent with the boundary conditions \(\psi_n(0) = 0\) and \(\lim_{\xi \rightarrow \infty} \psi_n(\xi) = 0\), \(\forall n\). It is noteworthy that if one takes \(\Lambda_0 = 0\) (the Hermitian limit), the wavefunctions and the spectrum turn out to be
\begin{equation}
\psi_n(\xi) \sim e^{- \frac{\eta  \xi^2}{4}} \xi^{\frac{1}{2} \big(\sqrt{\eta ^2+1}+1\big)} L_{n}^{\frac{\sqrt{\eta ^2+1}}{2}}\left(\frac{\eta \xi^2}{2}\right), \quad \quad E_n = \frac{8V_0}{\eta} \left(n+\frac{1}{2}+\frac{\sqrt{\eta ^2+1} - \eta}{4}\right),
\end{equation}
coinciding exactly with the results of \cite{iso1,iso2}. 

\section{Pseudo-Hermitian extensions for general potentials}
It should be remarked that one can generate other pseudo-Hermitian Hamiltonians via coupling the system in an arbitrary potential with an imaginary-valued gauge field. In the position representation, Eq. (\ref{HgencoupledtoA}) reads as 
\begin{equation}\label{Hpos}
H =  -\frac{1}{2m} \frac{d^2}{dx^2} + V(x) + \frac{A(x)^2}{2m} + \frac{i A(x)}{m}  \frac{d}{dx} + \frac{i A'(x) }{2m} . 
\end{equation}
Defining \(g(x) = e^{-i \int A(x) dx}\), we have 
\begin{eqnarray}
h &=& g(x) H g(x)^{-1} \nonumber \\
&=& -\frac{1}{2m} \frac{d^2}{dx^2} + V(x), \label{hexpressionpositionspace}
\end{eqnarray} which is just the corresponding Hermitian Hamiltonian. The above-mentioned analysis reveals that \(H\) and \(h\) share the same spectrum and the eigenfunctions of \(H\) can be determined if one can solve the time-independent Schr\"odinger equation for \(h\), which assumes a simpler form. Indeed, if \(\psi_h(x)\) be an eigenfunction of \(h\) with eigenvalue \(E\), then 
\begin{equation}
h \psi_h(x) = E \psi_h(x), 
\end{equation} which, upon substituting \(h = g(x) H g(x)^{-1}\) gives \(g(x) H (g(x)^{-1} \psi_h(x)) = E \psi_h(x)\). This directly implies that
\begin{equation}
 H (g(x)^{-1} \psi_h(x)) = E (g(x)^{-1} \psi_h(x)), 
\end{equation} meaning that \(\psi_H(x) = g(x)^{-1} \psi_h(x)\) is the corresponding eigenfunction of \(H\) with the same eigenvalue. As special cases, we can obtain the examples worked out so far in this paper. Let us look at another simple case in which the gauge field is an imaginary-valued constant, i.e., \(A(x) = - i \Delta\), \(\Delta \in \mathbb{R}\); the Hamiltonian then reads
\begin{equation}
H =  -\frac{1}{2m} \frac{d^2}{dx^2} +V(x) - \frac{\Delta^2}{2m} + \frac{ \Delta}{m} \frac{d}{dx} ,
\end{equation} while
\begin{equation}
h = -\frac{1}{2m} \frac{d^2}{dx^2} + V(x).
\end{equation} 
Notice that \(H\) is non-Hermitian and it is also not \(\mathcal{PT}\)-symmetric. However, \(H\) is still pseudo-Hermitian with \(h\) being its Hermitian equivalent. If \(V(x) =  \frac{m \omega^2 x^2}{2}\), the spectrum of \(h\) is obtained trivially as
\begin{equation}
E_n = \omega \bigg( n + \frac{1}{2} \bigg), \quad \quad n=0,1,2,\cdots,
\end{equation} which is real and equispaced. This example was presented earlier in \cite{example}, albeit in a different form. Thus, \(H\) also has the same spectrum whose reality is ensured by its pseudo-Hermiticity despite absence of \(\mathcal{PT}\)-symmetry. This demonstrates that \(\mathcal{PT}\)-symmetry is not a necessary condition to ensure the reality of the spectrum.

\section{Intertwined Hamiltonians}\label{susysec}
While several authors have studied supersymmetry and supersymmetric factorization in the context of non-Hermitian quantum systems (see for example, the works \cite{susynh1,susynh2,susynh3,susynh4}), let us now describe the supersymmetric factorization of the pseudo-Hermitian Hamiltonians considered in this paper. Consider some generic `supercharge-like' operators \(Q_1\) and \(Q_2\) such that (we shall take \(m=1\))
\begin{equation}\label{Qdef1}
Q_1 = \frac{1}{\sqrt{2}} \frac{d}{dx} + K(x) + W(x), \quad Q_2 = -\frac{1}{\sqrt{2}} \frac{d}{dx} - K(x) + W(x),
\end{equation} where \(K(x)\) is some real-valued function and \(W(x)\) is the superpotential. Notice that \(Q_1\) and \(Q_2\) are not Hermitian conjugates of one another. 
This gives
\begin{equation}
H_{\rm I} = Q_1 Q_2 = -\frac{1}{2} \frac{d^2}{dx^2} - \sqrt{2} K(x) \frac{d}{dx} + \frac{(W'(x) - K'(x))}{\sqrt{2}} + W(x)^2 - K(x)^2,
\end{equation}
\begin{equation}
H_{\rm II} = Q_2 Q_1 = -\frac{1}{2} \frac{d^2}{dx^2} - \sqrt{2} K(x) \frac{d}{dx} - \frac{(W'(x) + K'(x))}{\sqrt{2}} + W(x)^2 - K(x)^2,
\end{equation} which can be regarded as non-Hermitian superpartners with the non-Hermiticity originating from the fact that \(Q_1\) and \(Q_2\) are not Hermitian conjugates of one another. The Hamiltonians \(H_{\rm I}\) and \(H_{\rm II}\) become Hermitian if we set \(K(x) = 0\). Here \(W(x)\) can be taken to be the usual superpotential as in a Hermitian problem while \(K(x)\) is related to the imaginary-valued gauge field. Let us take two examples below. 

\subsection{Pseudo-Hermitian extension of harmonic oscillator}
For \(x \in \mathbb{R}\), taking \(W(x) = \frac{\omega x}{\sqrt{2}}\) and \(K(x) = - \frac{\lambda x}{\sqrt{2}}\), where \(\omega, \lambda >0\), it follows that the partner Hamiltonians turn out to be
\begin{eqnarray}
H_{\rm I} &=&  -\frac{1}{2} \frac{d^2}{dx^2} + \lambda x \frac{d}{dx} + \frac{(\omega^2 - \lambda^2) x^2}{2} + \frac{\lambda + \omega}{2},\\
H_{\rm II} &=&  -\frac{1}{2} \frac{d^2}{dx^2} + \lambda x \frac{d}{dx} + \frac{(\omega^2 - \lambda^2) x^2}{2} + \frac{\lambda - \omega}{2}. 
\end{eqnarray}
These are a pair of intertwined pseudo-Hermitian Hamiltonians with Hermitian counterparts that go as
\begin{eqnarray}
h_{\rm I} &=&  -\frac{1}{2} \frac{d^2}{dx^2} + \frac{\omega^2 x^2}{2}  + \frac{ \omega}{2} , \label{h1extension11} \\
h_{\rm II} &=&  -\frac{1}{2} \frac{d^2}{dx^2} + \frac{\omega^2 x^2}{2}  - \frac{ \omega}{2}.  \label{h2extension11}
\end{eqnarray}
Thus, they have the same spectrum but with different ground-state energies given as
\begin{equation}
E_{{\rm I},n} = \omega (n + 1), \quad \quad E_{{\rm II},n} = \omega n, \quad \quad n=0,1,2,\cdots. 
\end{equation}

\subsection{Pseudo-Hermitian extension of isotonic oscillator}
Let us now consider the case for which
\begin{equation}
W(x) = \frac{1}{\sqrt{2}} \bigg(\omega x + \frac{1}{x}\bigg), \quad \quad K(x) = -\frac{\lambda x}{\sqrt{2}}, \quad \quad x > 0.
\end{equation} Notice that the superpotential is singular at \(x = 0\) and one is restricting oneself to \(x > 0\). The superpartner Hamiltonians now read (defined for \(x > 0\))
\begin{eqnarray}
H_{\rm I} &=&  -\frac{1}{2} \frac{d^2}{dx^2} + \frac{(\omega^2 - \lambda^2) x^2}{2}  + \frac{\lambda}{2} \bigg(2 x \frac{d}{dx} +  1\bigg) + \frac{3 \omega}{2} , \label{H1extension} \\
H_{\rm II} &=&  -\frac{1}{2} \frac{d^2}{dx^2} + \frac{(\omega^2 - \lambda^2) x^2}{2} + \frac{1}{x^2}  + \frac{\lambda}{2} \bigg(2 x \frac{d}{dx} +  1\bigg) + \frac{ \omega}{2}.  \label{H2extension}
\end{eqnarray}
The form of \(H_{\rm II}\) above can be taken to dictate a pseudo-Hermitian extension of the isotonic oscillator. It is straightforward to show using Eq. (\ref{hexpressionpositionspace}) that the Hermitian counterparts are (for \(x > 0\))
\begin{eqnarray}
h_{\rm I} &=&  -\frac{1}{2} \frac{d^2}{dx^2} + \frac{\omega^2 x^2}{2}  + \frac{3 \omega}{2} , \label{h1extension} \\
h_{\rm II} &=&  -\frac{1}{2} \frac{d^2}{dx^2} + \frac{\omega^2 x^2}{2} + \frac{1}{x^2} + \frac{ \omega}{2}.  \label{h2extension}
\end{eqnarray}
The spectrum of \(h_{\rm I}\) corresponds to the levels of the half harmonic oscillator, i.e.,
\begin{eqnarray}\label{E1spectrumharm}
E_{{\rm I},n} = \omega \big(n + 2\big), \quad \quad n = 1,3,5,\cdots,
\end{eqnarray} or \(E_{{\rm I},n} = \omega \big(2n + 3) \) if \(n = 0,1,2,\cdots\). Following the references \cite{iso1,iso2}, one finds the spectrum of \(h_{\rm II}\) to be 
\begin{equation}\label{E2spectrumharm}
E_{{\rm II},n} =  \omega \big(2n + 3\big), \quad \quad n = 0,1,2,\cdots.
\end{equation}
Thus, the pseudo-Hermitian extension of the isotonic oscillator is the superpartner of the pseudo-Hermitian extension of the half harmonic oscillator, i.e., supersymmetry acts only on the wavefunctions of the harmonic oscillator that vanish at \(x = 0\) (see also, the older work \cite{susyiso1}).

\section{Closing remarks}\label{dsec}
In this paper, we have analyzed a certain quadratic but non-Hermitian extension of the quantum harmonic oscillator with the model being equivalent to the well-known Swanson oscillator for \(\Omega \geq 0\). We have analytically solved for the wavefunctions and spectrum in the position representation for \(\Omega \geq 0\). The Hamiltonian is pseudo-Hermitian with its Hermitian equivalent being the harmonic oscillator which facilitates its exact solution using methods familiar from the analysis of the (Hermitian) harmonic oscillator. We have also presented and analyzed a pseudo-Hermitian extension of the isotonic oscillator and have explored its supersymmetric factorization.

\section*{Acknowledgements} We are thankful to the anonymous referee for constructive comments. We thank Bijan Bagchi for many useful correspondences. AG would like to thank Miloslav Znojil and Anindya Ghose Choudhury for multiple enlightening discussions. The work of AG is supported by the Ministry of Education (MoE), Government of India in the form of a Prime Minister's Research Fellowship (ID: 1200454). AS would like to acknowledge the financial support from IIT Bhubaneswar in the form of an Institute Research Fellowship. We are grateful to the organizers (in particular, to David Berm\'udez Rosales) of the {\it Xth International Workshop on New Challenges in Quantum Mechanics: Graphene, Supersymmetry, and Mathematical Physics} for the opportunity to present this work. AG is grateful to the Czech Technical University in Prague for hospitality during the final stages of preparing the manuscript. 



\begin{thebibliography}{99}


   \bibitem{NH}
    Rotter I 2009
   	{\it J. Phys. A: Math. Theor.} \textbf{42} 153001
  
   \bibitem{NH0}
    Michishita Y and Peters R 2020
   	{\it Phys. Rev. Lett.} \textbf{124} 196401
  
  
    \bibitem{NH1}
    Holmes K, Rehman W, Malzard S and Graefe E M 2023
   	{\it Phys. Rev. Lett.} \textbf{130} 157202




 \bibitem{NH2}
    Graefe E M, H\"oning M and Korsch H J 2010
   	{\it J. Phys. A: Math. Theor.} \textbf{43} 075306
	
	 \bibitem{NH3}
    G\'omez-Le\'on \'A, Ramos T, Gonz\'alez-Tudela A and Porras D 2022
   	{\it Phys. Rev. A} \textbf{106} L011501


 \bibitem{NH4}
    Niu X, Li J, Wu S L and Yi X X 2023
   	{\it Phys. Rev. A} \textbf{108} 032214

	
	
	 \bibitem{PT1}
    Bender C M and Boettcher S 1998
   	{\it Phys. Rev. Lett.} \textbf{80} 5243
	
	 \bibitem{PT2}
    Bender C M, Boettcher S and Meisinger P N 1999
   	{\it J. Math. Phys.} \textbf{40} 2201

     \bibitem{PTZ}
 Znojil M 1999
 {\it Phys. Lett. A} \textbf{259} 220
 
  \bibitem{PTZ1}
 Znojil M 2020
 {\it Sci. Rep.} \textbf{10} 18523
 
   \bibitem{Bagchi1}
Bagchi B and Quesne C 2000
{\it Phys. Lett. A} \textbf{273} 285
 
  \bibitem{Bagchi2}
Bagchi B and Roychoudhury R 2000
{\it J. Phys. A: Math. Gen.} \textbf{33} L1

    \bibitem{BH}
    Bender C M and Hook D W 2023
   	{\it arXiv:2312.17386}


 \bibitem{PT3}
    El-Ganainy R, Makris K G, Khajavikhan M, Musslimani Z H, Rotter S and Christodoulides D N 2018
   	{\it Nat. Phys.} \textbf{14} 11
    
 \bibitem{MUS}
    Musslimani Z H, Makris K G, El-Ganainy R and Christodoulides D N 2008
{\it Phys. Rev. Lett.} \textbf{100} 030402


    \bibitem{COR}
 Correa F and Plyushchay M S 2012
 {\it Phys. Rev. D} \textbf{86} 085028
 
  \bibitem{Mich1}
Noble J H, Lubasch M and Jentschura U D 2013
{\it Eur. Phys. J. Plus} \textbf{128} 93
 
 
   \bibitem{Mich2}
 Noble J H, Lubasch M, Stevens J and Jentschura U D 2017
 {\it Comput. Phys. Commun.} \textbf{221} 304


 
\bibitem{example}
 da Provid\^encia J, Bebiano N and da Provid\^encia J P 2011
	{\it Braz. J. Phys.} \textbf{41} 78
 
 



 \bibitem{phot1}
  Feng L, El-Ganainy R and Ge L 2017
   	{\it Nat. Photonics} \textbf{11} 752


 \bibitem{phot2}
  El-Ganainy R, Khajavikhan M, Christodoulides D N and Ozdemir S K 2019
   	{\it Commun. Phys.} \textbf{2} 37
	
	

     \bibitem{WANG}
     Wang C, Fu Z, Mao W, Qie J, Stone A D and Yang L 2023
     {\it Adv. Opt. Photonics} \textbf{15} 442
     




	
	
 \bibitem{EPO4}
  Li A, Wei H, Cotrufo M, Chen W, Mann S, Ni X, Xu B, Chen J, Wang J, Fan S, Qiu C W, Al\`u A and Chen L 2023
   	{\it Nat. Nanotechnol.} \textbf{18} 706
	
	

	
	 \bibitem{EPAMO}
 Liang C, Tang Y, Xu A N and Liu Y C 2023
   	{\it Phys. Rev. Lett.} \textbf{130} 263601




\bibitem{deco2}Schlosshauer M 2005 
{\it Rev. Mod. Phys.} \textbf{76} 1267

\bibitem{decorev}Schlosshauer M 2019 
{\it Phys. Rep.} \textbf{831} 1

	
	\bibitem{mostafa}
   Mostafazadeh A 2002
   	{\it J. Math. Phys.} \textbf{43} 205
	
		\bibitem{jones}
    Jones H F 2005
   	{\it J. Phys. A: Math. Gen.} \textbf{38} 1741

	
		\bibitem{jakubsky}
  Jakubsk\'y V 2007
   	{\it Acta Polytech.} \textbf{47} 71
	
	\bibitem{mostafa2}
   Mostafazadeh A 2010
   	{\it Int. J. Geom. Methods Mod. Phys.} \textbf{07} 1191


	\bibitem{das}
   Das A 2011
   	{\it J. Phys.: Conf. Ser.} \textbf{287} 012002
	
	  \bibitem{znojiltd} Znojil M 2008
    {\it Phys. Rev. D}  \textbf{78} 085003
	

    \bibitem{fring} Fring A and Moussa M H Y 2016
    {\it Phys. Rev. A} \textbf{94} 042128
    
     \bibitem{ORTIZ} 
    Zelaya K and Rosas-Ortiz O 2021
   {\it Quantum Rep.} \textbf{3} 458

	



	
	 \bibitem{swanson1}
    Swanson M S 2004
   	{\it J. Math. Phys.} \textbf{45} 585
	

  
	 \bibitem{swanson2}
    Graefe E M, Korsch H J, Rush A and Schubert R 2015
   	{\it J. Phys. A: Math. Theor.} \textbf{48} 055301
	
	
	 \bibitem{swanson3}
    Bagchi B and Marquette I 2015
   	{\it Phys. Lett. A} \textbf{379} 1584

 
	
	\bibitem{RAM} 
Fern\'{a}ndez V, Ram\'irez R and Reboiro M 2022
{\it J. Phys. A: Math. Theor.} \textbf{55} 015303
	
	
	\bibitem{bagchiEP}
  Bagchi B, Ghosh R and Sen S 2022
   	{\it EPL} \textbf{137} 50004

	
	
	\bibitem{iso1}
Gol'dman I I and Krivchenkov V D 1961
{\it Problems in Quantum Mechanics}
(London: Pergamon Press) 


\bibitem{iso2}
Weissman Y and Jortner J 1979
{\it Phys. Lett. A} \textbf{70} 177

\bibitem{img} 
Hatano N and Nelson D R 1996
{\it Phys. Rev. Lett.} \textbf{77} 570

\bibitem{img1} 
Oztas Z and Candemir N 2019
{\it Phys. Lett. A} \textbf{383} 1821

\bibitem{img2} 
Wong S and Oh S S 2021
{\it Phys. Rev. Research} \textbf{3} 033042

\bibitem{img3} 
Midya B 2024
{\it Phys. Rev. A} \textbf{109} L061502

\bibitem{img4} 
Qi Y, Pi J, Wu Y, Lin H, Zheng C and Long G L 2024
	{\it Phys. Rev. B} \textbf{110} 075411


\bibitem{CV} 
Chalykh O A and Veselov A P 2005
{\it J. Nonlinear Math. Phys.} \textbf{12} 179

\bibitem{ISO4} 
Guha P and Ghose Choudhury A 2013 
{\it Rev. Math. Phys.} \textbf{25} 1330009


\bibitem{ISO6} 
Thirulogasanthar K and Saad N 2004
{\it J. Phys. A: Math. Gen.} \textbf{37} 4567

\bibitem{lagu}
Abramovitz M and Stegun I A 1965
{\it Handbook of Mathematical Functions: With Formulas, Graphs, and Mathematical Tables}
(New York: Dover Publications)





\bibitem{susynh1} 
Znojil M, Cannata F, Bagchi B and Roychoudhury R 2000 
{\it Phys. Lett. B} \textbf{483} 284

\bibitem{susynh2} 
Alexandre J, Ellis J and Millington P 2020 
{\it Phys. Rev. D} \textbf{101} 085015

\bibitem{susynh3} 
Bagarello F 2020
{\it Math. Phys. Anal. Geom.} \textbf{23} 28

\bibitem{susynh4} 
Znojil M 2020 
{\it Symmetry} \textbf{12} 892

\bibitem{susyiso1} 
Casahorr\'an J 1995 
{\it Physica A} \textbf{217} 429








\end{thebibliography}
\end{document}